\documentclass[
    ,final            
  ]
  {aipproc}

\def\etal{et al.~}
\layoutstyle{6x9}
\usepackage{amsmath}

\begin{document}

\title{Central engine afterglow of Gamma-ray Bursts}


\author{Yi-Zhong Fan$^{1,2,3}$, Tsvi Piran$^3$ and Da-Ming Wei$^{1,2}$} {
  address={$^1$ Purple Mountain Observatory, Chinese Academy of
Science, Nanjing 210008, China.\\
$^2$ National Astronomical Observatories, Chinese Academy of
Sciences, Beijing 100012, China.\\
$^3$ The Racah Inst. of Physics, Hebrew University, Jerusalem 91904,
Israel.} }

\begin{abstract}
Before 2004, nearly all GRB afterglow data could be understood in
the context of the external shocks model. This situation has changed
in the past two years, when it became clear that some afterglow
components should be attributed to the activity of the central
engine; i.e., the {\it central engine afterglow}. We review here the
afterglow emission that is directly related to the GRB central
engine. Such an interpretation proposed by Katz, Piran \& Sari,
peculiar in pre-{\it Swift} era,  has become generally accepted now.

\end{abstract}

\maketitle

\noindent%
\section{Two kinds of GRB afterglows}
In the context of standard fireball model of Gamma-ray Bursts
(GRBs), the prompt $\gamma-$ray emission is powered by internal
shocks and the afterglow emission arises due to external shocks (see
\cite{Piran04} for a review). In the pre-{\it Swift} era, most of
the afterglow data was collected hours after the prompt $\gamma-$ray
emission. These data was found to be consistent with the external
forward shock model, though at times energy injection, a wind medium
profile, or a structured/patchy jet were needed. We call the
emission relevant to the external shocks generated by the GRB
remnant as the ``{\it fireball afterglow}" or ``afterglow". An
alternative possibility for the production of the afterglow is a
continued activity of the central engine (i.e., the {\it central
engine afterglow}) via either the ``internal shocks" or ``magnetic
dissipation". This idea has been put forward by Katz, Piran \& Sari
already in 1998, shortly after the discovery of the afterglow of GRB
970228. However, the agreement of the predictions of the external
shock afterglow model \cite{Sari98} with most subsequent
multi-wavelength afterglows observation strongly disfavors the {\it
central engine afterglow} interpretation.

\begin{figure}[h]
\includegraphics*[width=.9\textwidth]{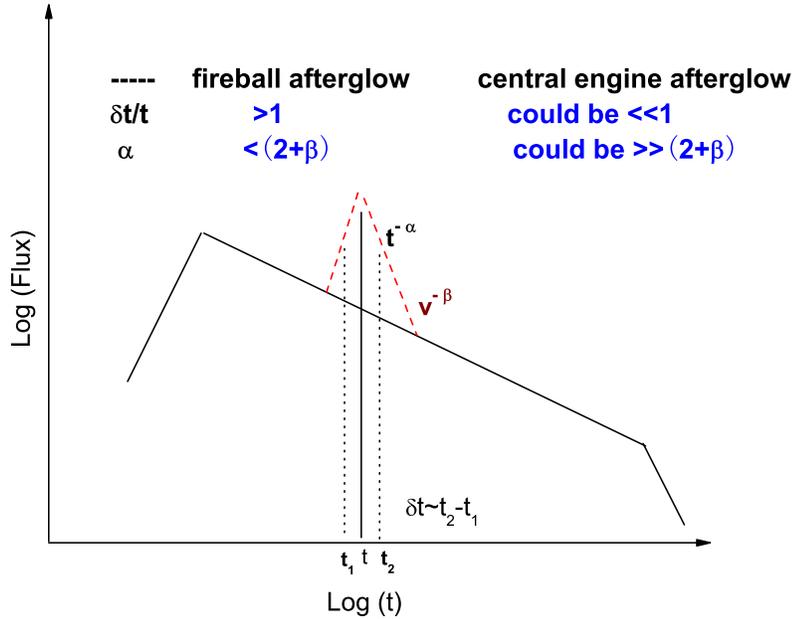}
\caption{Two constraints that help us to distinguish between the
{\it central engine afterglow} and the {\it fireball afterglow}.}
\label{fig:Illustration}
\end{figure}

One disadvantage of the central engine afterglow model is its lack
of predictive power. The fireball afterglow model, instead, has
predicted smooth light curves and in particular the intrinsic
relation between the flux in different bands (e.g., \cite{Sari98})
as well as between the spectral slops and the temporal decay. In
2003, it was already clear that in the case of a fireball afterglow,
the variability timescale of the emission $\delta t$ divided by the
occurrence time $t$ has to be in order of $1$ or larger \cite{NP03}.
A highly relevant constraint is that the decline of the fireball
afterglow emission can not be steeper than $t^{-(2+\beta)}$, (where
$\beta$ is the spectral index) unless the edge of the GRB ejecta is
visible. This is because the GRB outflow is curving and emission
from high latitude (relative to the observer) will reach us at later
times and give rise to a decline shallower than $t^{-(2+\beta)}$
\cite{Fenimore96,KP00}. These limitations do not apply, of course to
a central engine afterglow (see Fig.\ref{fig:Illustration}).

\section{Possible candidates of central engine afterglow}
{\bf Very early rapid X-ray decline.} If the central engine does not
turn off abruptly, its weaker and weaker activity  will give rise to
rapidly decaying emission \cite{FW05}. This model can account for
some very early rapid X-ray declines identified by the {\it Swift}
satellite \cite{Bart05}, in particular those decay with time more
slowly than $t^{-(2+\beta)}$, for which the high latitude emission
interpretation \cite{Zhang06} is invalid.

{\bf X-ray/optical flares.} In 1998, Piro et al. reported a
late-time outburst of the X-ray afterglow of GRB 970508 from
BeppoSAX. The authors attributed such an outburst to the re-activity
of the central engine. However, refreshed shocks (produced by slowly
moving matter that was ejected more or less simultaneously with the
faster moving one during the onset of the prompt emission) can
reproduce the multi-wavelength outburst as well \cite{PMR98}. So it
is not a good {\it central engine afterglow} candidate. In 2005,
Piro et al. \cite{Piro05} published an analysis of the X-ray data of
GRB 011121, in which two X-ray flares after the prompt $\gamma-$ray
emission are evident. They have  interpreted the X-ray flares, in
particular the first one, as the onset of the forward shock
emission. Fan \& Wei \cite{FW05} applied the {\it decline argument}
to these flares and suggested  a central engine origin. Many other
possibilities like a reverse shock, a density jump, a patchy jet,
energy injection and a refreshed shock have been convincingly ruled
out \cite{Zhang06}.

The XRT onboard {\it Swift} confirmed Piro et al.'s discovery
\cite{Burrows05}. By now X-ray flares have been well detected in
$\sim 40\%$ Swift GRBs. Most of these flares violate the two
constraints of the fireball afterglow (e.g., \cite{CG07}). Though
the physical processes powering these delayed X-ray bursts are not
clear yet, it is most likley that they are related to a re-activity
of the central engine, i.e., they are {\it central engine
afterglow}.

In a few bursts, optical flares have been detected (e.g.,
\cite{Boer06}) and those may also have a central engine origin
\cite{Wei06}. But in most bursts, the very early X-ray and optical
emission are not correlated (e.g., \cite{Molinari07}). This is a
natural result if the X-ray emission is dominated by the central
engine emission component while the central engine optical emission
has been suppressed by the synchrotron self-absorption, as suggested
by Fan \& Wei \cite{FW05}.

{\bf Power-law decaying X-rays.} The temporal behavior of the
flaring X-rays are quite similar to that of the prompt
$\gamma-$rays. It is thus reasonable that the flares and the prompt
GRB have a common origin. However, it is somewhat surprising to note
that even some power-law decaying X-ray light curve might be a {\it
central engine afterglow}. A good example is the afterglow of GRB
060218. The inconsistence of the X-ray afterglow flux with the radio
afterglow flux and the very steep XRT spectra support here the
central engine afterglow hypothesis \cite{FPX06,Sode06}. So far,
there are two more candidates GRB 060607A \cite{Molinari07} and GRB
070110 \cite{Troja07}. Both events are distinguished by a very sharp
X-ray drop in the afterglow phase, which is inconsistent with the
fireball afterglow interpretation \cite{JF07,Troja07}. This is in
particular the case for GRB 070110 because the optical data
simultaneous with the X-ray drop does not steepen at all. For GRB
060607A, the constraint is less tight because the late-time optical
light curve is unavailable.

The long lasting X-ray afterglow flat segment before the sharp X-ray
drop (both its luminosity and timescale) is well consistent with the
emission powered by the magnetic dissipation of a millisecond
magnetar wind, as suggested by Gao \& Fan \cite{GF06}. Of course,
additional independent signature, like a high linear polarization,
is needed before the magnetar wind dissipation model can be
accepted.

\section{Discussion}
In contrast to what was believed in the pre-{\it Swift} era, it is
evident now that the role of the central engine has to be taken into
account when interpreting  many {\it Swift} GRB afterglows. The ``ad
hoc" hypothesis made in Katz et al. \cite{Katz98}  has been well
confirmed. The chromatic behavior of the afterglow suggests that the
afterglow in X-ray and in lower energy bands may have a different
origin \cite{FW05}. This can be easily understood if the synchrotron
self-absorption is strong enough to suppress the central engine
optical emission but not the X-rays \cite{FW05}. Recently,
Ghisellini et al. \cite{GG07} adopted this idea to interpret the
chromatic break of the early optical and X-ray data of some {\it
Swift} GRBs \cite{FP06} and argued these shallowly decaying X-ray
emission are {\it central engine afterglow}.

The fruitful {\it Swift} early afterglow observations open a new
window to reveal the behavior of the central engine. The continued
activity of the GRB central engine is now a well-established fact.
But the underlying physical processes are less clear. Among the
various models put forward (see \cite{Z06} for a review) fallback
accretion onto the nascent black hole may be the most natural one.
Other models involve the magnetic activity of the central engine
\cite{FZP05,Dai06,GF06}. Such models might be tested by a
polarimetry as in magnetic energy dissipation scenarios one can
expect the emission to be highly polarized. Alternatively if the
shocks powering the X-ray {\it central engine afterglow} (the flares
or some power-law decaying light curves) are not significantly
magnetized the synchrotron-self Compton (SSC) emission will peak in
the GeV energy band and it may be strong enough to be detected by
the upcoming GLAST satellite (e.g., \cite{Fan07}). Such emission
might have been already observed in the long-lasting shallowly
decaying GeV afterglow emission of GRB 940217 \cite{Hurley94}. This
might have been be the SSC component of the central engine X-ray
flat segment as that detected in GRB 070110. If it is the case, a
rapid GeV emission drop simultaneous with that in X-ray band, will
be present. Our interpretation thus can be tested in the near future
directly by the cooperation of {\it Swift} and GLAST.\\

This work is supported by US-Israel BSF. TP acknowledges the support
of Schwartzmann University Chair. YZF and DMW are also supported by
the National Science Foundation (grant 10673034) of China and by a
special grant of Chinese Academy of Sciences.


\begin{thebibliography}{}
\bibitem{Bart05} Barthelmy, S. D., et al., 2005,
Nature, 438, 994
\bibitem{Boer06} Bo\"er, M., Atteia, J. L., Damerdji, Y., Gendre, B., Klotz,
A., \& Stratta, G. 2006, ApJ, 638, L71
\bibitem{Burrows05} Burrows D. N., et al., 2005, Science, 309, 1833
\bibitem{CG07} Chincarini G., et al., 2007
(astro-ph/0702371)
\bibitem{Dai06} {Dai, Z. G., Wang, X. Y., Wu, X. F., Zhang,
B.}, 2006, {Science}, {311}, {1127}
\bibitem{FP06} Fan Y. Z., Piran T., 2006, MNRAS, 369, 197
\bibitem{Fan07} Fan Y. Z., Piran T., Rarayan R., Wei D. M., 2007, MNRAS, in press (arXiv:
astro-ph/0704.2063)
\bibitem{FPX06} Fan Y. Z., Piran T., Xu D., 2006, JCAP, 0609, 013
\bibitem{FW05} Fan Y. Z., Wei D. M., 2005, MNRAS, 364, L42
\bibitem{FZP05} Fan Y. Z., Zhang B., Proga D., 2005, ApJ, 635, L129
\bibitem{Fenimore96} Fenimore E. E., Madras C. D., Nayakshin S., 1996,
ApJ, 473, 998
\bibitem{GF06} Gao W. H., Fan Y. Z., 2006, Chin. J. Astron. Astrophys,
              6, 513
\bibitem{GG07} Ghisellini G., Ghirlanda G., Nava L., Firmani C.,
2007, ApJ, 658, L75
\bibitem{Hurley94} Hurley K., et al., 1994, Nature, 372, 652
\bibitem{JF07} Jin Z. P., Fan Y. Z., 2007, MNRAS, 378, 1043
\bibitem{Katz98} Katz J. I., Piran T., Sari R., 1998, Phys. Rev.
Lett., 80, 1580
\bibitem{KP00} Kumar P., Panaitescu A., 2000, ApJ, 541, L51
\bibitem{Molinari07} Molinari E. \etal, 2007, A\&A, 469, L13
\bibitem{NP03} Nakar E., Piran T., 2003, ApJ, 598, 400
\bibitem{PMR98} {Panaitescu, A., Meszaros, P., Rees, M. J.}, 1998,
{ApJ}, 503, 314
\bibitem{Piran04} Piran T., 2004, Rev. Mod. Phys., 76, 1143
\bibitem{Piro98} Piro L., et al., 1998, A\&A, 331, L41
\bibitem{Piro05} Piro L., et al., 2005, ApJ, 623, 314
\bibitem{Sari98} Sari R., Piran T., Narayan R. 1998, ApJ, 497, L17
\bibitem{Sode06} Soderberg A.  et al.,  2006, Nature, 442, 1014
\bibitem{Troja07} Troja E. \etal, 2007,
ApJ, 665, 599
\bibitem{Wei06} Wei D. M., Yan T., Fan Y. Z., 2006, ApJ, 636, L69
\bibitem{Z06} Zhang B., 2006, AIPC, 838, 392
\bibitem{Zhang06} Zhang B., Fan Y. Z., Dyks J., Kobayashi S., M\'esz\'aros P.,
Burrows D. N., Nousek J. A., Gehrels N.  2006, ApJ, 642, 354
\end{thebibliography}
\end{document}